\documentclass[letter]{aa}  
\usepackage{graphicx}
\catcode`\@=11
\def\gtsima{\ifmmode{\mathrel{\mathpalette\@versim>}}
    \else{$\mathrel{\mathpalette\@versim>}$}\fi}
\def\ltsima{\ifmmode{\mathrel{\mathpalette\@versim<}}
    \else{$\mathrel{\mathpalette\@versim<}$}\fi}
\def\@versim#1#2{\lower 2.9truept \vbox{\baselineskip 0pt \lineskip 
    0.5truept \ialign{$\m@th#1\hfil##\hfil$\crcr#2\crcr\sim\crcr}}}
\catcode`\@=12
\usepackage{txfonts}
\begin{document}
   \title{GMASS Ultradeep Spectroscopy of Galaxies at $z\sim2$. III: The emergence of the color bimodality at $z\sim2$}
   \titlerunning{Color bimodality at 0$<z<$3}

   \author{P. Cassata \inst{1}, A. Cimatti \inst{2}, J. Kurk \inst{3},
     G. Rodighiero \inst{4}, L. Pozzetti \inst{5}, M. Bolzonella
     \inst{5}, E. Daddi \inst{6}, M. Mignoli \inst{5}, S. Berta
     \inst{7}, M. Dickinson \inst{8}, A. Franceschini \inst{4},
     C. Halliday \inst{9}, A. Renzini \inst{10}, P. Rosati \inst{11}
     \and G. Zamorani \inst{5}} \authorrunning{P. Cassata et al.}
     \offprints{P. Cassata}

   \institute{Laboratoire d'Astrophysique de Marseille, OAMP, UMR6110,
     CNRS-Universit\'e de Provence Aix-Marseille I, BP8, F-13376 Marseille
     Cedex 12, France
              \email{paolo.cassata@oamp.fr}
         \and
         Dipartimento di Astronomia, Universit\`a di Bologna, via Ranzani 1,
         I-40127, Bologna, Italy
         \and
         Max-Planck-Institut f\"ur Astronomie, K\"onigstul 17,
         D-69117, Heidelberg
         \and
         Dipartimento di Astronomia, Universit\`a di Padova, Vicolo
         dell'Osservatorio 2, I-35122, Padova
         \and
         INAF-Osservatorio Astronomico di Bologna, Via Ranzani 1, I-40127,
         Bologna
         \and
         CEA-Saclay, DSM/DAPNIA/Service d'Astrophysique, F-91191 Gif-sur-Yvette
         Cedex, France
	 \and
	 Max Planck Institut f\"ur Extraterrestrische Physik, Postfach 1312, 
	 85741 Garching bei M\"unchen, Germany
	 \and
	 NOAO-Tucson, 950 North Cherry Avenue, Tucson, AZ 85719, USA
	 \and
	 INAF-Osservatorio Astrofisico di Arcetri, Largo E. Fermi 5, I-50125,
	 Firenze, Italy
	 \and
	 INAF-Osservatorio Astronomico di Padova, Vicolo dell'Osservatorio 5,
	 I-35122, Padova, Italy
         \and
         European Southern Observatory, Karl-Schwarzschild-Strasse 2, D-85748,
         Garching bei M\"unchen, Germany
}

   \date{Received .....; accepted .....}

  \abstract 
{} 
  {The aim of this work is to study the evolution of the rest
  frame color distribution of galaxies with the redshift, in
  particular in the critical interval $1.4<z<3$.}
  {We combine ultradeep spectroscopy from the GMASS project ({\it Galaxy
    Mass Assembly ultradeep Spectroscopic Survey}) with GOODS
    multi-band photometry (from optical to mid-infrared) to study a
    sample of 1021 galaxies up to m(4.5$\mu$m)=23.}  
  {We find that the distribution of galaxies in the $(U-B)$ color vs
    stellar mass plane is bimodal up to at least redshift $z=2$. We
    define a mass complete sample of galaxies residing on the
    red-sequence, selecting objects with $\log(M/M_{\odot})>10.1$, and
    we study their morphological and spectro-photometric
    properties. We show that the contribution to this sample of
    early-type galaxies, defined as galaxies with a spheroidal
    morphology and no star formation, decreases from 60-70\% at
    $z<$0.5 down to $\sim$50\% at redshift $z=2$. At $z>2$ we still
    find red galaxies in the mass complete sample, even if the
    bimodality is not seen any more. About 25\% of these red galaxies
    at $z>2$ are passively evolving, with the bulk of their stars
    formed at redshift $z>3$.}
{}


   \maketitle
%

\section{Introduction}
Galaxies in the local universe show a bimodal color distribution
(Strateva~et~al.~2001, Hogg~et~al.~2002, Blanton~et~al.~2003). Even if
color bimodality has been observed and studied also at higher redshift
(Bell~et~al.~2004, Weiner~et~al.~2005, up to $z\sim1$;
Franzetti~et~al.~2006, Cirasuolo~et~al.~2007, up to $z\sim1.5$;
Giallongo~et~al.~2005, up to $z\sim2$), no study until now has combined
spectroscopic coverage and morphological analysis at $z>1$.
The color bimodality suggests a different mechanism of 
evolution for galaxies lying on the two sequences (Menci~et~al.~2005;
Scarlata~et~al.~2007; De~Lucia~et~al.~2007): hence, it is extremely
interesting to evaluate the epoch when the color bimodality was built
up.

Many authors in the literature use red sequence galaxies to constrain
the evolution of the early-type population (Bell~et~al.~2004a;
Faber~et~al.~2007). Although the red peak is known to be dominated by
early-type galaxies with old passive stellar populations, a
contamination of star forming galaxies with colors reddened by dust,
or by early-type spirals is also present (Bell~et~al.~2004b;
Cassata~et~al.~2007; Franzetti~et~al.~2007; Scarlata~et~al.~2007).
Here we use data about spectral properties, spectral
energy distributions (SED), morphologies and mid-IR emission that we
collected for the GMASS sample to explore the content of the red
sequence as a function of the redshift.
Throughout the paper, magnitudes are in the AB system, and we adopt
$H_0$=70~km~s$^{-1}$~Mpc$^{-1}$, $\Omega_m$=0.3 and
$\Omega_{\Lambda}$=0.7.


\section{The sample}
GMASS {\it(``Galaxy Mass Assembly ultra-deep Spectroscopic Survey'')}
is a project based on an ESO VLT Large Program. For an exhaustive
description of the survey see Kurk~et~al.~(2008a, in preparation),
Cimatti~et~al.~(2008) and Halliday~et~al.~(2008). 

The sample is extracted in the $4.5\mu$m public image obtained with
{\it Spitzer Space Telescope} + IRAC down to a limiting magnitude of
$m_{4.5}=$23.0, and contains 1021 galaxies in a field of
6.8'x6.8'. The selection at $4.5\mu$m is more sensitive to stellar
mass and less affected by dust extinction than optical bands. Moreover
this selection produces a ``negative'' k-correction at $z>1.4$, as the
peak of the stellar spectral energy distribution enters the $4.5\mu$m
band at this redshift.

   \begin{figure}
   \centering
   \includegraphics[width=8.cm,bb=15 52 393 757]{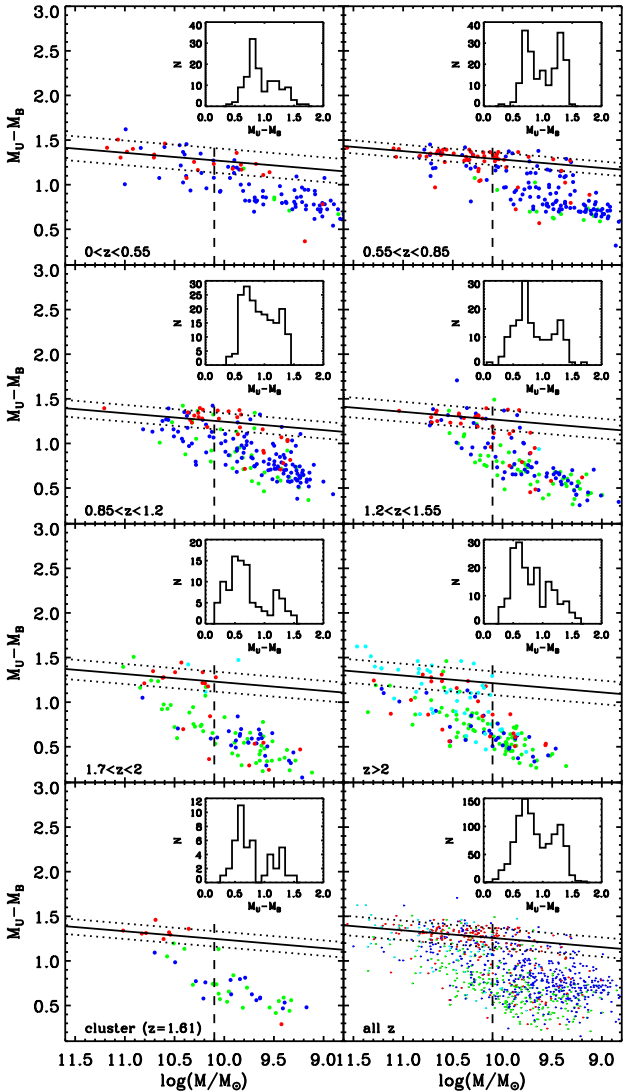}
   \caption{Rest-frame $(U-B)$ color versus stellar mass in six
     redshift intervals. A separate panel is dedicated to the
     forming cluster at $\langle z\rangle=1.61$. The diagonal
     continuous line indicates fit to the red sequence, as the dotted
     ones show the scatter around the best fit, for the different
     redshift bins. The dashed line indicates 10.1 $M/M_{\odot}$, the
     mass limit at which the red sequence is complete at any redshift
     in our sample. Color-coded symbols indicate galaxies in different
     morphological classes: red, blue, green and cyan symbols
     represent respectively early-types, spirals, irregulars and
     undetected objects. In each panel the histogram of colors is also
     reported. The bottom-right panel reports the color-mass
     relation for all the galaxies, regardless of the redshift.}
   \label{color_mass}%
    \end{figure}

With the aim of studying the redshift range $1.5 < z < 3$, the
epoch when the crucial processes of massive galaxy formation took
place, ultra deep spectroscopy was carried out using FORS2 on
galaxies pre-selected with a cut in photometric redshift of
$z_{phot}>$1.4. The spectroscopic coverage was complemented with
available literature redshifts, reaching a completeness of about
50\%. For the rest of the population, accurate photometric redshifts
are available (Kurk~et~al.~2008a). The median of the galaxy redshift
distribution is $\langle z\rangle=1.2$, with 190 objects with
spectroscopic redshift $z>1.4$, mostly coming from GMASS spectroscopy.
There are several spikes in the redshift distribution, the most
dominant being at $\langle z\rangle=1.61$, discussed in a parallel
paper (Kurk~et~al.~2008b).

The photometric SEDs were derived for all the galaxies in the sample
using the public images available in the GOODS-South area in 11 bands:
optical (ACS-HST, $BVIz$, Giavalisco~et~al.~2004), near-infrared (ESO
VLT+ISAAC, $JHK_S$) and mid-infrared (Spitzer+IRAC, 3.6$\mu$m,
4.5$\mu$m, 5.6$\mu$m and 8$\mu$m).  We used the synthetic
spectra of Maraston~(2005; M05) to fit these SEDs, adopting a Kroupa
IMF, limiting the fit to $\lambda_{rest}<2.5\mu$m. We use
exponentially declining star formation histories,
with timescales spanning from 0.1 to 30 Gyr, plus the case of constant
star formation rate. Models with ages between 0.1 Gyr and the age of
the Universe at the redshift of each galaxy are retained in the
best-fit procedure. Extinction is a free parameter in the
optimization, using the extinction curve of
Calzetti~et~al.~(2000). The fitting procedure minimizes the $\chi^2$,
and the best fit model gives an estimate for the age, the $e-$folding
time of the SFR $\tau$, the extinction $A_V$ and the stellar mass. At
the same time, absolute magnitudes in Johnson $UBV$ bands are derived.
The dataset is complemented with the Spitzer-MIPS data publicy
available for the GOODS-South/GMASS region, to check for possible
activity signs in the 24$\mu$m data.

The high resolution imaging provided by ACS-HST allowed an accurate
visual classification of all the galaxies in the sample. The analysis
has been performed independently by two of us (PC and GR) on the ACS
band closest to the rest-frame $B-$band.  The classification scheme is
based on 4 classes:
1.  spheroidal galaxies (ellipticals, S0 and compact objects);
2. spirals; 3. irregular galaxies; 4. undetected in the optical bands
(and thus not classifiable).
On the basis of this analysis, we classified 198 spheroidal galaxies,
496 spirals and 269 irregulars, while 58 objects are undetected in the
optical bands. 

At redshift $z>1.2$, where the ACS $z$-band maps the blue UV
light in the rest-frame, surface brightness dimming and
morphological k-correction effects may be important. However, the
separation between early- and late-types should remain robust, as
--though fainter in the UV-- ellipticals remain symmetrical. Spirals
instead can appear more late type, as red bulges get fainter, the
surface brightness of disks dimms, but knots of star formation
brighten in the UV.


\section{Color bimodality}
Precise redshift measurements are extremely important to reduce
uncertanties in the SED fitting procedure, to derive robust
absolute magnitudes, stellar masses, and other SED parameters. The
GMASS spectroscopic coverage in the range $1.4<z<2.5$ allows on one
hand a proper calibration of the photometric redshifts, and, on the
other hand, provides us with high-quality fits of the spectral energy
distribution for each galaxy. The optimal measure of the rest-frame
magnitudes allows us to check for the bimodality in colors in a redshift
interval that has not been much explored until now.
In Figure~\ref{color_mass} we report the rest-frame $(U-B)$ color
versus stellar mass for different redshift bins, as well as for the
global population. Each redshift bin contains roughly the same number
of objects.
In each panel a small inset shows the color distribution at that
redshift. The forming cluster at $\langle z\rangle=1.61$ is excluded
from the bins at $z>1.2$, but we dedicated a separate panel to it. The
quality of our sample allows us to claim that the color distribution,
looking at both the $U-B$ histograms and the $(U-B)$ vs $M/M_{\odot}$
plane, is bimodal at least up to $z=2$, as well as for the global
population.  We note that galaxies show also a sort of bimodality in
mass, with red galaxies being more massive of the blue ones. This is
partly due to a selection effect: at a given redshift, the
completeness limit is higher for red than blue galaxies. Conversely,
on the other side of the diagram, the lack of massive blue galaxies is
a real effect.

To characterize and define the red sequence, first we select
objects redder than the valley in the global color distribution,
namely (U-B)=1.1.  Then, we fit the global color-mass distribution of
these galaxies with a straight line (U-B)=$a$+$b$(log(M/M$_{\odot})$,
obtaining for the slope a value $b$=0.0943.  Finally, we use this
slope to fit the red sequence in each redshift bin, leaving only the
intercept $a$ as a free parameter. The result of this procedure is
overplotted on the data in Fig.~\ref{color_mass}. Interestingly, the
intercept $a$ does not evolve significantly with redshift: we measure
a $\Delta(U-B)$=-0.2 between $z\sim$0.5 and $z\sim$2.5. This is
consistent with the evolution expected for a galaxy formed in a
instantaneous burst at redshift $z=5$.

In Figure \ref{color_mass} we color coded galaxies of different
morphologies. We find that 70\% of the spheroidal galaxies reside on
the red sequence. The remaining 30\% show blue colors. On average,
red spheroidal galaxies are more massive than blue ones (the median of
the mass distribution is $2\times10^{10}$ and $5\times10^{9}$ $M_{\odot}$
respectively for red and blue).  On the other hand, 77\% of late type
galaxies have blue colors, with the remaining 23\% lying on the red
sequence.  
At redshifts $z>2$ many objects have colors z-K$>2$, being very faint
or undetected up to the ACS $z-$band. For these no morphological
analysis could be made.

\section{Red sequence composition}
In general, galaxies can have a red color because they have old
stellar populations, or because their star-formation activity is
obscured by dust that reddens their colors. Hence, the red sequence is
well known to be a mix of galaxies with different properties
(Bell~et~al.~2004b; Cassata~et~al.~2007; Scarlata~et~al.~2007;
Franzetti~et~al.~2007). Here we have the opportunity to explore the
morphological, photometric and spectroscopic properties of red
galaxies up to redshift $z\sim3$.  

To separate the blue and red population in each redshift bin, we use
the best fit to the red sequence and add an offset of $\Delta m=-0.15$
magnitudes to end up on the valley between red and blue galaxies. This
offset is comparable to the scatter of the objects around the fit to
the red sequence.  To allow a comparison between galaxies at different
redshifts, we define a mass complete sample, selecting only
galaxies on the red-sequence having log$M/M_{\odot}>$10.1. This is
roughly the smallest mass for red galaxies at redshift $z\sim3$ in our
sample. With this criterion, we select a sample of 197 galaxies. The
aim of this section is to study the properties of these galaxies, by
combining morphological information and SED analysis, including also
$24\mu$m data.

   \begin{figure}
   \centering
   \includegraphics[width=6.cm]{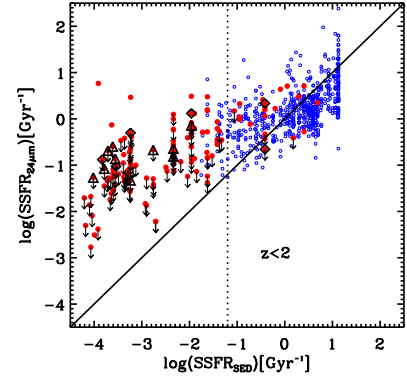}
   \caption{Comparison between Specific Star Formation Rates (SSFR)
     coming from SED fitting and from 24$\mu$m luminosity, for
     galaxies at $z<2$. Red points represent galaxies lying on the red
     sequence, having log(M/M$_{\odot})>10.1$ (the mass completeness
     limit at any redshift), while blue empty circles show galaxies in
     the blue cloud. Empty triangles are galaxies with passive
     spectra, empty diamonds are galaxies with passive spectra but
     some signs of emission lines. Downward arrows identify galaxies
     for which the 24$\mu$m luminosity is just an upper limit, as
     the they have $S_{24\mu m}<25\mu$Jy.
}
   \label{ssfr}%
    \end{figure}

In Figure \ref{ssfr} we study the star formation properties of
galaxies in this sample, comparing the Specific Star Formation Rates
(the star formation rate for unit mass, SSFR) estimates coming from
SED fitting and from 24$\mu$m photometry. We plot only galaxies up to
$z=2$, because beyond this limit the data show to much scatter.  We
report also blue galaxies for comparison. The total infrared
luminosity (8-1000 $\mu$m) was computed by rescaling various star
forming templates to the 24$\mu$m flux of each source (see
Daddi~et~al.~2007 for details). Then, this total luminosity was
converted into a Star Formation Rate using the relation given by
Kennicutt~(1998). 
We use 25$\mu$Jy as a conservative detection limit for 24$\mu$m
sources in GOODS (corresponding roughly to $S/N<2$), reporting as
upper limits the values of the SSFR below this limit.

The agreement between the two SSFR estimates up to $z=2$ is quite
good, in particular for blue galaxies, even if a large scatter is
present. The two measurements still correlate for red massive
galaxies, considering also that the $24\mu$m SSFR provides an upper
limit for the bulk of them. The large majority of the mass complete
red sample has $\log$(SSFR$_{SED})<-1.2$ and no reliable detection
at 24$\mu$m.
The quite good correlation between the two SSFR and the agreement with
spectroscopic properties ensure that this diagram can be used to
identify galaxies which have very low levels of SSFR and are likely to
be the ones with the lowest or no star formation. In particular, we
define passive those galaxies having $\log$(SSFR$_{SED})<-1.2$: at
this rate, a galaxy would take about 15 Gyr to double its mass.
To be more conservative, we excluded galaxies having a
reddening $A_V>$1 and those with $\log$(SSFR$_{24\mu m})>-1$ and
$S_{24\mu m}>25\mu$Jy (that are reliable $24\mu$m sources).  We
checked the position on the diagram of the 16 galaxies in the
mass complete sample having passive GMASS spectra (see
Cimatti~et~al.~2008 for the 13 at $z>1.4$): 14 of them occupy the
``passive'' region of the diagram, while 2, that also show some
emission lines in their spectra, are in the ``active'' region.

At $z>2$ the scatter about the relation becomes significant and we
therefore decided not to plot galaxies with $z>2$ on Fig.~\ref{ssfr};
this scatter is probably due to increased uncertanties on
the 24$\mu$m luminosity and/or to an increasing number of dusty star
forming objects. In this range we decided to use the same limits
to define passive galaxies, which exclude objects with large 24$\mu$m
luminosities.
Finally, among the 197 galaxies belonging to the mass complete sample,
103 ($\sim50\%$) could be considered passive according to these
criteria.

It is interesting to combine the results of the morphological
classification with those of the analysis of the star formation
activity, for our mass complete sample of galaxies on the red
sequence.  We find that all the objects classified as spheroidals are
actually passive galaxies. This means that at we do not find, at any
redshift, red sequence massive ellipticals with a secondary episode of
star formation. On the other hand, not all the galaxies that are
passive according to the criteria mentioned above are morphologically
spheroidals. In fact, about 20\% of the 103 passive galaxies on the
red sequence are morphologically late-types. The large majority of
them are spirals dominated by a bright bulge.

\begin{figure}
   \centering \includegraphics[width=6.cm]{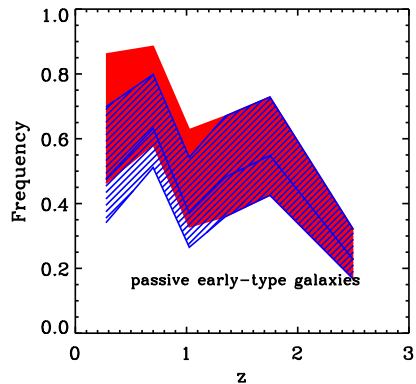}
   \caption{Contribution to the massive part of the red sequence
     (log(M/M$_{\odot})>10.1$) as a function of the redshift, for
     early-type galaxies with little or no signs of star
     formation. Red strip includes bulge dominated spirals, while blue
     dashed one includes pure spheroidal galaxies.  }
   \label{red_seq}%
    \end{figure}

We summarize these results in Fig.~\ref{red_seq}, where we show the
contribution to the massive part of the red sequence
($\log(M/M_{\odot}>10.1$) of galaxies that show little or no signs of
star formation
($\log$(SSFR$_{SED})<-1.2$~\&$~A_V<1$~\&$~\log$(SSFR$_{24\mu m})<-1$)
and have an early-type morphology. Since at $z>2$ we do not have a
morphological classification for all the galaxies, at that redshift we
combine SED and 24$\mu$m information to define galaxies with low
SSFR. In the literature, authors either include or exclude bulge
dominated spirals as a part of the early-type family, we therefore
show here both cases. It can be noted that the trend of the two cases
is similar, with the contribution of passive early-type galaxies
mildly (or not at all) decreasing with the redshift from $z\sim$0 to
$z\sim2$. The fluctuations in the frequency could be due to
cosmic variance: i.e, the lack of massive galaxies on the red sequence
in the bin around $z\sim$1 may imply that our survey samples an
underdense region of the Universe at that redshift (see the
photometric redshift histogram in Kurk~et~al.~2008a).  The results at
redshift $z<1.5$ are in good agreement with previous results:
Franzetti~et~al.~(2007) find an increasing contribution of star
forming galaxies in the red sequence between $z\sim$0 and
$z\sim1.2$. Moreover, Scarlata~et~al.~(2007) found that the fraction
of morphologically early-type galaxies (including bulge dominated
spirals) in a sample of photometrically selected ETGs decreases from
$\sim$60\% at $z=0.3$ to $\sim45$\% at $z=0.9$ .  Renzini~(2006)
reported that at z$\sim 0$ 58\% of the red galaxies are morphologically
early-type, similar to our findings. At z$\sim 0.7$
Bell~et~al.~(2004b) and Cassata~et~al.~(2007) find respectively 75\%
and 66\% of early-types contributing to the red sequence: these
results can be easily conciliated with ours considering that the
former include Sa galaxies, while the latter consider pure
ellipticals/S0.

Here we reach higher redshift. We show that even at $1.5<z<2$ the
color distribution is bimodal, with the massive part of the red
sequence dominated by passive spheroidal galaxies
(Fig.~\ref{color_mass}\&\ref{red_seq}). Even though at $z>2$
bimodality in the color distribution is no longer evident, many
galaxies do show red colors and masses above
$\log(M/M_{\odot})=10.1$. We find that $\sim25\%$ of them have low or
no signs of star formation (compared to ~60\% at $z<2$), having
$\log$(SSFR$_{SED})<-1.2$ \& $\log$(SSFR$_{24\mu m})<-1$ \&
$~A_V<1$. Half of these red and massive galaxies have a spheroidal
morphology, even if the remaining half can not be morphologically
classified, because they are barely detected or undetected in the ACS
$z-$band. Nevertheless, all of them have $BzK$ colors compatible with
being passive (Daddi~et~al.~2004). Combining their ages (greater than
0.5 Gyr) to their redshift, we can conclude that this population of
massive galaxies must have formed the bulk of their stars during a
burst at redshift $z>3$. It is plausible to argue that the bimodality
at $z>2$ vanishes as a result of the decreasing number of passive red
galaxies and of the consequent increase of dusty star forming objects.

Finally, we emphasize that, even at $z<1$ and more significantly at
$z>1$, the fraction of massive late-type galaxies
lying on the sequence is significant, and evolves with redshift. Thus,
it could be misleading to identify red-sequence with passive old
galaxies. This must be taken into account when red sequence galaxies
are used to constrain the evolution of the luminosity and mass
function for the passive early-type population, as in
Bell~et~al.~(2004a), Scarlata~et~al.~(2007) and Faber~et~al.~(2007).

\section{Conclusions}
     We study the rest-frame $(U-B)$ color distribution for a sample of
     1021 galaxies selected at $4.5\mu$m. We took advantage of the
     ACS/HST high resolution imaging and multiwavelength coverage spanning
     from ACS bands to Spitzer IRAC mid-infrared and 24$\mu$m to
     characterize morphologies, colors and SED properties of galaxies in
     the sample.

   \begin{enumerate}
   \item We find that the classical color bimodality is preserved up
     to at least $z=2$. This can be seen both looking at the galaxies
     in the (U-B) vs Mass plane, and at the $(U-B)$ distribution.
     The presence in the sample of a forming cluster at $\langle
     z\rangle=1.61$ does not enhance the bimodality at $1.5<z<2$.
   \item The massive part of red sequence ($\log(M/M_{\odot})>10.1$)
     is a mix of early- and late-type galaxies at all redshifts. The
     contribution of early-type galaxies to the red sequence 
     is about 60-70\% at $z\sim0.5$ and about 50\% at $z=2$, depending
     on the inclusion or exclusion of bulge dominated passive spirals
     in the early-type class.
   \item Even if the color bimodality vanishes at redshift $z>2$,
     still there are red galaxies with $\log(M/M_{\odot})>10.1$.
     About 25\% of these are passively evolving, with ages $>$0.5
     Gyr. This implies that we find a rich population of massive
     galaxies with redshift of formation $z>3$.
   \end{enumerate}

\end{document}